\documentclass[10pt, conference]{IEEEtran}
\IEEEoverridecommandlockouts
\usepackage{soul}
\usepackage{cite}
\usepackage{amsmath,amssymb,amsfonts}
\usepackage{algorithmic}
\usepackage{graphicx}
\usepackage{textcomp}
\usepackage{url}
\usepackage{booktabs}
\usepackage{caption}
\usepackage{subcaption}

\usepackage{comment}

\usepackage{multirow}
\usepackage{booktabs}
\usepackage{graphicx}

\usepackage[colorinlistoftodos]{todonotes}

\def\BibTeX{{\rm B\kern-.05em{\sc i\kern-.025em b}\kern-.08em
    T\kern-.1667em\lower.7ex\hbox{E}\kern-.125emX}}
    
\begin{document}

\title{Personal Data Protection in Smart Home Activity Monitoring for Digital Health:\\A Case Study}

\author{
    \IEEEauthorblockN{Claudio Bettini, Azin Moradbeikie, Gabriele Civitarese}
    \IEEEauthorblockA{
        \textit{EveryWare Lab, Dept. of Computer Science} \\
        \textit{University of Milan}, Milan, Italy\\
        \{claudio.bettini, azin.moradbeikie, gabriele.civitarese\}@unimi.it
    }
}

\maketitle

\begin{abstract}
Researchers in pervasive computing have worked for decades on sensor-based human activity recognition (HAR). Among the digital health applications, the recognition of activities of daily living (ADL) in smart home environments enables the identification of behavioral changes that clinicians consider as a digital bio-marker of early stages of cognitive decline. The real deployment of sensor-based HAR systems in the homes of elderly subjects poses several challenges, with privacy and ethical concerns being major ones. This paper reports our experience applying privacy by design principles to develop and deploy one of these systems.
\end{abstract}

\begin{IEEEkeywords}
smart home, human activity recognition, privacy
\end{IEEEkeywords}

\textcolor{red}{\textit{This paper has been accepted and presented at the TELMED '25 workshop. It will appear in the 2025 IEEE International Conference on Pervasive Computing and Communications Workshops and other Affiliated Events (PerCom Workshops) proceedings.}}

\section{Introduction}
Human Activity Recognition (HAR) is a very active research area, with significant progress partly due to new sensing technologies and new AI methods \cite{Chen21-survey,arrotta2022dexar}.
Sensor-based HAR enables several healthcare applications, including the early detection of cognitive decline. 

Using remote, unobtrusive, and continuous monitoring to diagnose early stages of cognitive decline, its phases, and its possible progression to dementia through the analysis of Activities of Daily Living (ADLs) is a challenge of clinical interest~\cite{grammatikopoulou2024assessing}.
Indeed, several projects have investigated the sensor-based recognition of human activities with real deployments in smart home targeting the specific application of monitoring cognitive decline as revealed by behavioral changes of the monitored subjects \cite{cook2012casas,riboni2016smartfaber, palermo2023tihm}. 
The lack of a reliable privacy-preserving infrastructure is, among other challenges, one of the reasons why there have not been large-scale deployments of these systems. Privacy is by itself a research area in mobile and pervasive systems \cite{bettini2015privacy}. Real systems should be built with privacy by design principles, adopting end-to-end state-of-the-art solutions, complying with the new regulations, and reassuring the data subjects.  

In this paper, we report our experience in designing and deploying a privacy-aware infrastructure as part of a new project in this area. While the number of monitored homes is again pretty limited, this is a \emph{pilot} for a possibly large-scale regional or national deployment.

\section{The SERENADE project}
The SERENADE project aims to develop a telemedicine solution for the detection, through intelligent analysis of long-term sensor data, of behavioral changes suggesting the evolution of Mild Cognitive Impairment (MCI). 
MCI is a pre-stage of dementia where the subject, even if presenting cognitive dysfunctions on one or multiple domains, still keeps their daily functionality. For some individuals, MCI may progress to dementia. The clinical goal of this project is to identify unobtrusive digital biomarkers to monitor functional and symptomatic changes in MCI patients living alone, in order to capture the early transition from MCI to dementia.

The project team includes experts in sensor-based activity recognition, clinicians (neurologists and neuropsychologists), and technicians managing the sensing infrastructure. This multidisciplinary team identified the key behaviors that can be monitor through sensors: nutrition, personal hygiene, sleep, therapy adherence, mobility, and cognition. Specifically, SERENADE leverages unobtrusive sensing devices such as environmental sensors (e.g., magnetic, PIR, and plug sensors), a smart sleep analyzer, a tablet to administer periodical cognitive tests, and a smartwatch. Activities of Daily Living (ADLs), like cooking and eating, are recognized using XAI methods on the stream of sensor data. Detected ADLs are then analyzed to infer long-term behavioral changes possibly linked to cognitive decline. These changes and corresponding explanations are inspected by the clinicians via a dedicated web dashboard during the clinical visits to support their diagnosis.

SERENADE considers two patient cohorts: one with MCI due to neurodegeneration, likely progressing to cognitive decline, and another with MCI not due to neurodegeneration, typically stable. 
Overall, we plan to recruit $15$ subjects for each group. At the time of writing, we deployed our system in the home of $18$ patients.

\section{Privacy Model}
In this section, we describe the privacy model by identifying the role of each entity in the data workflow, specifying the type of data exposed to each entity, and analyzing the possible threats.

In the following, we consider a privacy threat the acquisition of identified sensitive data by unauthorized entities. In addition to usual personal data, we focus on the protection of sensitive data related to the health of the subjects as well as to their activities and habits in daily life, when at home or outside. By \emph{identified} we mean that the data is associated with the identity of a specific subject.

\subsection{Data subjects, Data controller and  data processors}

\subsubsection{Data subject}
In Serenade, the \emph{data subjects} are the patients selected by the hospital, that will be monitored during the pilot. In the following, we will refer to them simply as \emph{subjects}. 


\subsubsection{Data controller}
The data controller is the hospital (HOS) that organizes the recruitment of the participating subjects among its patients. 

\subsubsection{Data processor}
The data processors in Serenade are the following:
The main data processor is our lab\footnote{Formally is Università degli Studi di Milano.} (UniMi) with a team of computer scientists that designs algorithms, performs data analysis, and offers a dashboard to visualize results for clinicians. 
A second data processor is the Infrastructure Installation Team (IIT)  which is in charge of installing the necessary devices (mini-PC, tablet, smartwatch, sensor devices, etc.) in the subjects’ homes and intervening during the study to fix possible malfunctions.
A third data processor is the Infrastructure Monitoring Team (IMT) which configures the devices to be installed and monitors the infrastructure to ensure that sensor data is correctly acquired and stored as well as transmitted to the central repository for future analysis. IMT should not be able to analyze data. They are just responsible for making sure of the correct configuration and operation of devices.
A fourth data processor is the COD20 Telemedicine technical team (TelMed) which will provide secure storage of the data, authentication and access control mechanisms, and a platform for interfacing with legacy systems.

Other (external) data processors are:
\begin{itemize}
    \item the company that manufactures the smartwatch and sleep analyzer systems and that receives through a smartphone app, stores, and preprocesses in its cloud the raw data obtained by these devices.
    \item the company that manufactures the smart toothbrush and that receives through a smartphone app, stores and preprocesses in its cloud the raw data obtained by this device.
    \item the company that acquires and processes the outdoor movements of the subject by its geolocation services activated on the subject’s smartphone.
\end{itemize}

\subsection{Data control}
Data processors have different requirements in terms of data access: 
\begin{itemize}
    \item UniMi needs to receive all sensing data, but it does not need to know the subject's identity. However, to correctly interpret sensor data, it may need some general information about the subject and the home environment. For example, it should know about the positioning of sensors in the apartment, if there are any pets, and if other individuals are living with or regularly visiting the subject.
    \item The IIT, to perform installation and maintenance interventions, needs to know the subject’s identity, home address, and phone number. It does not need to see the data acquired by the sensing infrastructure.
    \item The IMT operates as the system administrator of the sensing infrastructure, but it does not need to know the subject’s identity, nor any other information that may reveal the identity, as, for example, the location of the home.
\end{itemize}

\subsection{Type of Data}

For the sake of the study, for each participating subject, the data shown in Table~\ref{tab:D0}, called D0,need to be known to IIT.


\begin{table}[]
\caption{D0.}
\label{tab:D0}
\resizebox{\columnwidth}{!}{%
\begin{tabular}{|l|l|l|l|l|l|l|}
\hline
\begin{tabular}[c]{@{}l@{}}First \\ name\end{tabular} & \begin{tabular}[c]{@{}l@{}}Last \\ name\end{tabular} & Age & Gender & \begin{tabular}[c]{@{}l@{}}Complete \\ address\end{tabular} & \begin{tabular}[c]{@{}l@{}}Contact names + \\ phones (possibly \\ family members \\ or caregivers)\end{tabular} & \begin{tabular}[c]{@{}l@{}}General notes \\ (e.g., living with \\ caregiver, pet, type \\ of apartment, etc.)\end{tabular} \\ \hline
\end{tabular}%
}
\end{table}

While the hospital will have additional identifying information as well as complete medical records, this information needs to be shared with IIT, the team that is in charge of installing and maintaining the sensing infrastructure at the patient’s home.  This data is also sufficient for the medical personnel to link the subject to the rest of their medical data.
The IIT may also add information to the “General Notes” after the visits and installation in the apartment. For example, a map of the apartment with the position and type of the installed sensors will be added. 

The data acquired by the infrastructure can be divided into the following groups: 

D1: Data acquired through the Smartwatch, Sleep analyzer, and Smart Toothbrush is shown in Table~\ref{tab:D1} and referred to as D1.


\begin{table}[]
\caption{D1.}
\label{tab:D1}
\resizebox{\columnwidth}{!}{%
\begin{tabular}{@{}|l|l|l|l|l|l|@{}}
\hline
Daily steps & \begin{tabular}[c]{@{}l@{}}Daily Avg \\ heart rate\end{tabular} & \begin{tabular}[c]{@{}l@{}}Daily sleep \\ duration and \\ weekly sleep \\ cycle\end{tabular} & Sleep quality & \begin{tabular}[c]{@{}l@{}}Night breathing \\ quality (apnea, \\ snoring)\end{tabular} & \begin{tabular}[c]{@{}l@{}}Daily \\ brushing \\ time and \\ duration\end{tabular} \\ \hline
\end{tabular}%
}
\end{table}

D2: Data acquired by the Digital Assistant Cognitive Evaluation is shown in Table~\ref{tab:D2} and referred to as D2.


\begin{table}[]
\caption{D2}
\label{tab:D2}
\resizebox{\columnwidth}{!}{%
\begin{tabular}{@{}|l|l|l|@{}}
\hline
\begin{tabular}[c]{@{}l@{}}Report with answers to a 
\\ digital Mini-Mental State\\ test (monthly)\end{tabular} & \begin{tabular}[c]{@{}l@{}}Report with answers to a 
\\ digital  cognitive test (weekly)\end{tabular} & \begin{tabular}[c]{@{}l@{}}Data about the 
compliance  \\ 
of the subject in taking 
\\
the test (with date/time)\end{tabular} \\ \hline
\end{tabular}%
}
\end{table}

D3: Data acquired by the Environmental Sensors (time series) is shown in Table~\ref{tab:D3} and referred to as D3.


\begin{table}[]
\caption{D3}
\label{tab:D3}
\resizebox{\columnwidth}{!}{%
\begin{tabular}{@{}|l|l|l|l|@{}}
\hline
Presence in a room & \begin{tabular}[c]{@{}l@{}}Use of some appliances \\ (turn on/off)\end{tabular} & \begin{tabular}[c]{@{}l@{}}temp \& humidity \\ (kitchen, bathroom)\end{tabular} & \begin{tabular}[c]{@{}l@{}}Opening/closing of \\ doors/cabinets/drawers\end{tabular} \\ \hline
\end{tabular}%
}
\end{table}

D4: Data acquired by the Google Location History service through the subject’s smartphone is shown in Table~\ref{tab:D4} and referred to as D4.

\begin{table}[]
\caption{D4}
\label{tab:D4}
\resizebox{\columnwidth}{!}{%
\begin{tabular}{|l|l|l|l|l|l|l|l|}
\hline
\multicolumn{3}{|c|}{Visited Place} & \multicolumn{4}{c|}{Activity Segment} \\ \hline
\begin{tabular}[c]{@{}l@{}}Location\\ (lat, lon)\end{tabular} & \begin{tabular}[c]{@{}l@{}}Start\\ timestamp\end{tabular} & \begin{tabular}[c]{@{}l@{}}End\\ timestamp\end{tabular} & \begin{tabular}[c]{@{}l@{}}Start\\ location\\ (lat, lon)\end{tabular} & \begin{tabular}[c]{@{}l@{}}Start\\ timestamp\end{tabular} & \begin{tabular}[c]{@{}l@{}}End\\ location\\ (lat, lon)\end{tabular} & \begin{tabular}[c]{@{}l@{}}End\\ timestamp\end{tabular} \\ \hline
\end{tabular}%
}
\end{table}

\subsection{Entities and authorizations}

Considering the analysis on data control and type of data we can more precisely define which entities should be authorized to access which data.
The subject explicitly authorizes HOS to obtain all the sensitive data as well as the results of the data analysis for the following purposes:  a) better monitoring of the health conditions and possible early intervention if needed, and b) medical research related to neurological disorders. 
The IIT should be authorized to access D0 since it has to directly interact 
with the subject for installing and maintaining the infrastructure. However, it does not need to acquire any information about the sensitive data acquired by the infrastructure.
%
The IMT is not authorized to obtain D0 (i.e., data that can identify the subject)  but should be authorized to access the system side of the infrastructure for the only purpose of monitoring its correct operation. For monitoring it may be exposed to some sensed data and in this case, the data should not identify the subject. It is not allowed to store or process any of the data nor to see the results of data analysis.  
UniMi does not need any information in D0 to perform data analysis, hence it is not authorized to access it, except for some general information like the map with sensor positioning, but it will need access to all the data acquired through the infrastructure at each home location. Hence, it should be authorized to obtain and process sensitive data D1, D2, and D3, but not to associate the data or the analysis results to the specific subject. 

The external companies involved in Serenade monitoring impose a registration to their services to provide their data analysis. When installing the necessary app and registering, they get the subject’s name and email; Hence, in principle, they can associate the sensitive data sent to their cloud by their devices with the subject’s identity. They are authorized by the subject that explicitly provides consent also based on the security and privacy guarantees that these entities offer in handling and using this data, including compliance with the data protection regulations (e.g., EU GDPR). 

\subsection{Identifiers, quasi-identifiers, and related privacy risks}
\label{sec:QI}
A crucial step of any privacy analysis is to determine which data may reasonably act as a quasi-identifier.
This is data that, when linked with external information, may identify the subject or significantly restrict the candidate subjects \cite{bettini2008anonymous}. 
This task is also related to the \emph{adversary model} since this model determines the type of external information that may be accessible and the adversary's inference abilities (e.g., in terms of computational resources).
The adversaries considered in Serenade can be divided into internal adversaries (all the data processors) and external adversaries. 
We first consider what data could act as quasi-identifiers for internal adversaries.

While most of the data in D0 identifies the subject (some attributes are explicit identifiers and others are quasi-identifiers), the data included in D1, D2, and D3 even if combined is very unlikely to act as a quasi-identifier in Serenade, considering the external information that is available to internal adversaries. Indeed, while some data patterns may be unique to a given subject, UniMi, which is analyzing this data, has no reasonable way to get from external sources an association between these patterns and the subject’s identity. A different situation concerns D4, since the subject’s trajectories are geo-localized and may be used to infer the subject’s home address among other re-identifying locations and patterns. They are considered quasi-identifiers since they could reveal the subject’s identity (e.g., because the subject is the only elderly living at that address) [3]. 

Another potential quasi-identifier is the IP address of the home gateway that is known by the IMT (it is used by the IMT to connect with the Home gateway) and may be also obtained by UniMi for the connection through which pseudonymized data is provided. However, the internet connection is provided through SIM cards registered by UniMi to a mobile operator or WiFi. Hence there is no association between IP and the subject stored by the mobile provider, the IP is dynamic, and geolocation will only provide an imprecise position in a highly populated area.  The association between the SIM card and the subject is only available to the IIT that inserts the SIM card into the tablet before installing it in the subject’s home. We conclude that there is a negligible probability that unauthorized entities participating in the protocol may re-identify the subjects based on the IP address of their home gateway.

Regarding external potential attackers, by adopting state-of-the-art security measures to protect data in transit and data at rest we believe there is a relatively low risk of a privacy violation. For the same reason, the external companies involved as entities in the project as well as the IIT have no reasonable way to obtain sensitive data that they are not authorized to get.

\section{Privacy-Preserving Solution}

\subsection{Dataflow architecture and protocols}
\begin{figure}[h!]
    \centering
    \includegraphics[width=\columnwidth]{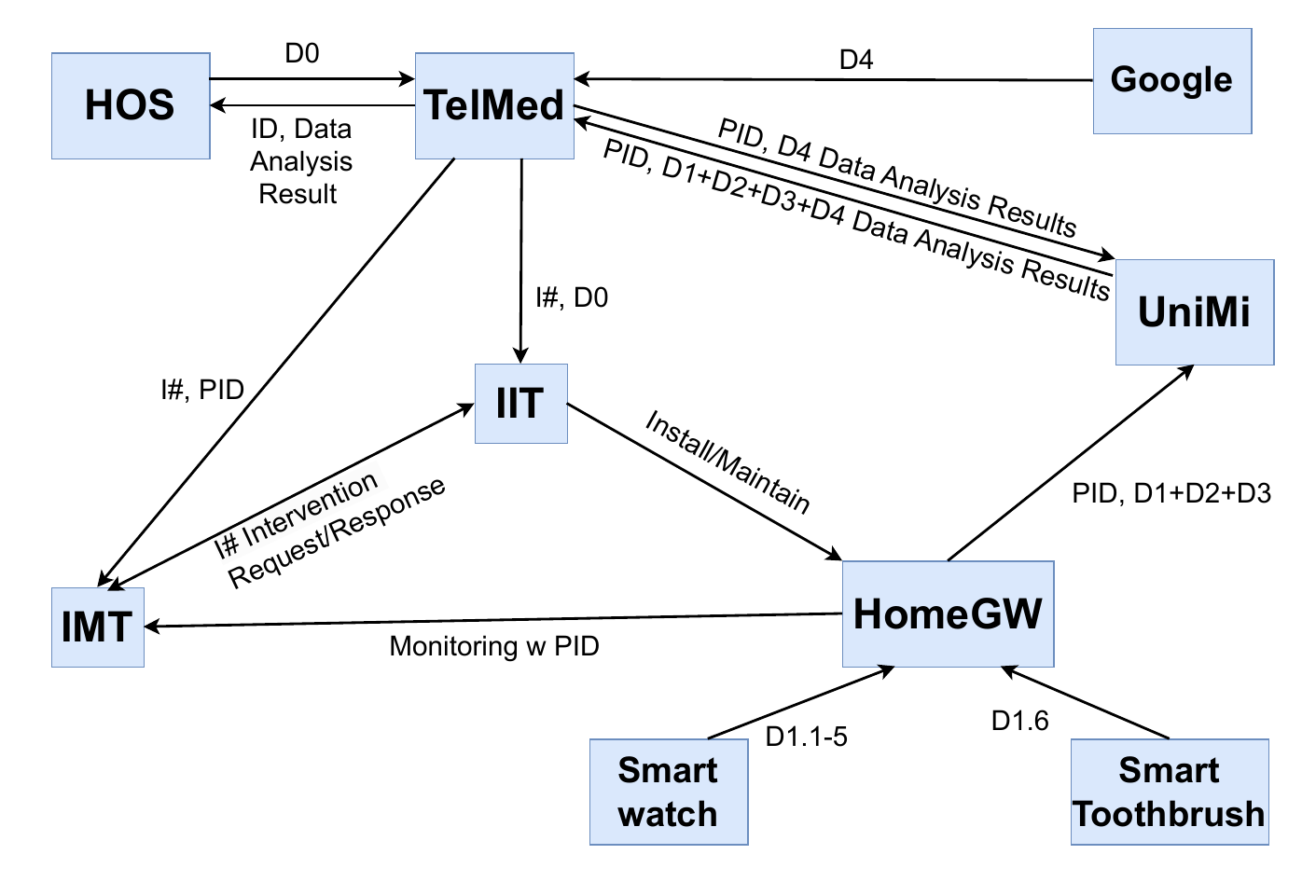}
    \caption{The dataflow architecture}
    \label{fig:dataflow}
\end{figure}

Figure~\ref{fig:dataflow} shows the data workflow model with each edge in the graph reporting the type of data transferred in the communication between the involved entities. Note that this drawing ignores the TelMed entity since it is considered a trusted party for the HOS.

We illustrate the dataflow considering first the \emph{initialization protocol} that defines what happens when a new subject is enrolled in the study. 
The process starts with the registration of the new subject's personal data (D0) by HOS. The system generates two codes: a pseudonym (PID) and an installation number (I\#). PID will be used instead of explicit identifiers to avoid re-identification of sensitive data. I\# will be only used in the communications between IIT and IMT to refer to the same home infrastructure. The association between the PID and the subject's identity is securely stored. Then, an "installation request" ticket including the information about D0 and I\# is created for IIT. A second message is sent from HOS to IMT communicating both PID and I\#.

IIT eventually processes the "installation request" ticket, by scheduling a visit to the subject's apartment. During the visit, the IIT plans the devices' installation and uploads some general notes that will be useful in interpreting sensor data (apartment map with sensor positioning and settings). Finally, it acknowledges the completion of this step via the ticket (a message to IMT with the I\# identifier).

IMT reacts to this message by preparing the infrastructure (e.g. binding the necessary sensing devices with the home gateway (HomeGW), configured with the PID corresponding to I\#). Then it notifies the IIT via the ticket. The IIT will pick up the devices, schedule the installation with the subject, and complete the setup. It then notifies the IMT. In the last step, the IMT confirms the system's functionality by connecting with the home gateway and verifying that data acquisition works as intended. Finally, the IMT terminates the initialization protocol by closing the installation ticket.

Suppose some issue with data acquisition occurs during the monitoring period. In that case, the IMT sends an intervention request (with I\#) to the IIT which will plan and conduct the intervention, notifying the IMT when solved.

All the data obtained in the home (D1+D2+D3) is stored in a local database on HomeGW and is periodically transferred to the data analysts (UniMi), associated with the subject's PID. 

As a separate process, the outdoor mobility data D4 obtained by Google is stored on a trusted server. A data analysis algorithm automatically processes this data, returning results of interest to clinicians but not geo-localized, so they should not act as quasi-identifiers. The results are transferred to UniMi associated with the PID for possible correlations with other data.

The clinicians of HOS receive from UniMi all the data analysis results associated with the subject's PID. 
%
%
The mapping between the PID and the data that HOS uses to identify the subject will be used by HOS to visualize the data analysis results on a dashboard for the specific subject possibly joined with the subject’s medical records.
It is worth mentioning that the retention of sensitive data at the HomeGW should be limited not only for storage limitation but more importantly for privacy reasons since it is possibly exposed to IMT. The UniMi database will keep the data for the whole duration of the study. After the study, the data will be anonymized by deleting the association between PID and subject and by performing a further analysis about the risk of re-associating the data with a subject.

\subsection{Data protection techniques}
This subsection outlines the techniques implemented to protect data in tables D0, D1, D2, and D3. The adopted technique for protecting D4 is presented in the next subsection.

The main privacy-preserving technique adopted in this pilot is \emph{pseudonymization} and its effectiveness strongly depends on the proper identification of identifiers and quasi-identifiers as presented above. A second principle that we follow is the \emph{separation of duties}. We assign different tasks to different entities, each one with a partial visibility of the data. To enforce this principle an authorization system based on access control must be implemented. This also enables strict control of the data flow between the different data processors. In more detail, these are the main techniques adopted in SERENADE.

\begin{itemize}
    \item Pseudonymization:
    We apply pseudonymization by removing identifying and quasi-identifying data and associating sensitive data with a code
    that is unique to each subject but it is very unlikely to be useful to re-associate sensitive data with the subject's identity. 
 There are different methods to properly generate pseudo-identifiers~\cite{doi/10.2824/860099, UKICO2022}. 
 Our implemented system uses AES-256 symmetric encryption to achieve pseudonymization. The encryption method takes the identifier data and a nonce as input and generates a corresponding PID for each subject. 
    \item Authorization: The system implements authorization to verify users and their permissions~\cite{barkadehi2018authentication}.  
%
    \item Access control: The system employs role-based access control, defining distinct roles for each of the entities involved and assigning specific permissions to each role~\cite{sandhu1998role}. Access is restricted to the data necessary for each role's tasks and responsibilities.
    \item Logging: all accesses to the system are logged with timestamp, type of access, and identifier of the entity that made the access. 
    
\end{itemize}

To prevent attacks from external adversaries, in addition to these techniques, standard security measures for protecting intrusion in the home local network (including IOT devices) and for protecting data in transit and data at rest should be adopted. 
In SERENADE all data is encrypted when transferred among the participating entities using state-of-the-art secure Web communication protocols. Particularly sensitive data like the association between PID and the subject's identity or the geo-localized data are stored in encrypted tables in the DBMS. Mechanisms like Transparent Data Encryption (TDE) provide an efficient solution by automatically encrypting and decrypting data stored in the database~\cite{sidorov2015transparent}.

    

\subsection{Privacy Preserving Analysis of D4}
\subsubsection{The Process of Collecting D4} 
Location and trajectory data (D4) is collected through the subject’s smartphone and recorded by Google’s location service~\cite{zheng2015trajectory}. For this purpose, the subject’s phone, with location-enabled services, continuously tracks their location and records location coordinates, timestamps, and activity types (e.g., walking, driving). Then, Google periodically compiles this data into time series objects, which include places visited (specific locations the subject stayed for a while) and activity segments (movements between visiting locations). At the end of each month, these time series objects are automatically uploaded to the trusted server as a JSON-formatted file.

\subsubsection{Objectives of the outdoor mobility analysis}

The primary goal that clinicians expect by monitoring activities outside the home is to identify deviations from normal habits. In this project, clinicians are interested in the following: 
\begin{itemize}
    \item Repeated deviations from the usual time spent outside the home: Unusually long or short periods spent away from home. We are interested in duration-per-day and event counting per day
    \item Changes in the places the subject frequently visits (e.g., the subject stops going to the usual grocery store and starts going to a closer one).
    \item Changes in the route taken to reach or come back from a frequently visited place (e.g., unusual longer route). 
    \item Identifying repeated "wandering" behavior: apparently irrational trajectories that suggest the subject may be lost or experiencing difficulty in navigating. 
\end{itemize}

\subsubsection{D4 Data Protection}
As observed in Section~\ref{sec:QI}, the geo-localised data in D4 can act as a quasi-identifier. Hence, it cannot be exposed to the data analysts (UniMi). 
In addition to store this data in an encrypted table, we designed algorithms that identify the changes in outside activities as requested by the clinicians without returning any (quasi-)identifying data in the output. Ideally, the geo-localized data should be protected also when \emph{in use}, by running these algorithms in a TEE (Trusted Execution Environment). 


\begin{figure*}[]
    \centering
    \includegraphics*[width=1\linewidth]{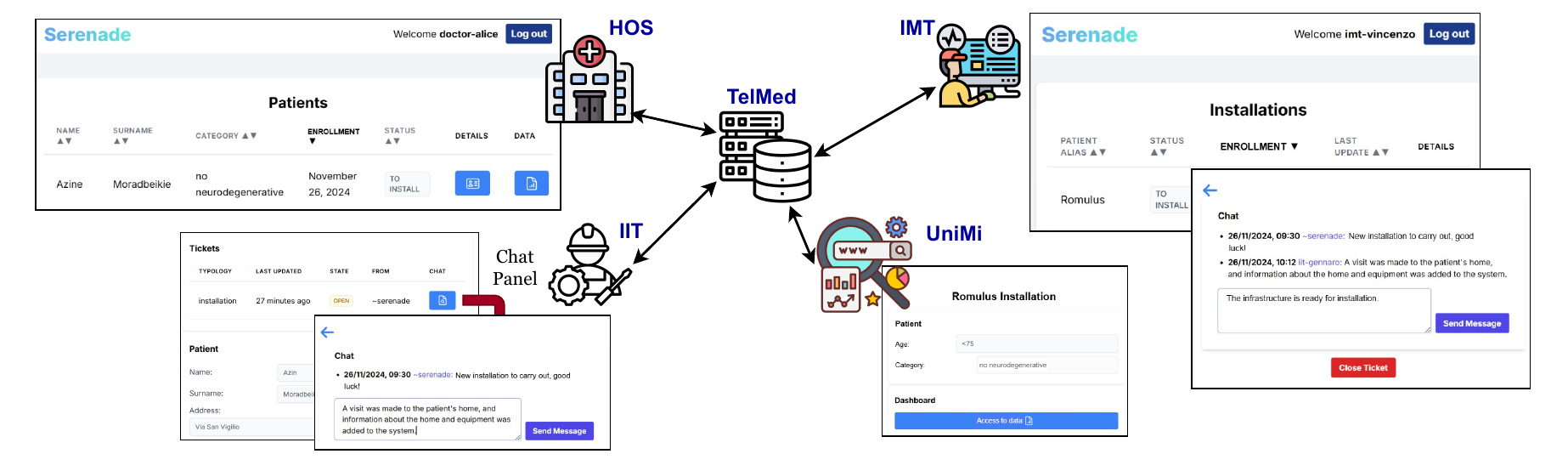}
    \caption{Some interfaces of the SERENADE application}
    \label{fig:application}
\end{figure*}

\subsection{Compliance with data protection regulations}
The project pilot has been approved by the reference IRB (ethical committee) of the hospital recruiting the subjects. The approval was based on a detailed description of the data being acquired, the data analysis being performed, the use of the data, the forms for informed consent that the participating subjects had to sign, and the system's implementation following the privacy-by-design approach to embed privacy principles directly into its architecture and processes.
\begin{itemize}
    \item Lawfulness, fairness, and transparency: patients provide explicit consent after being informed of the purpose of data collection, the processing methods, and how their information will be utilized. This ensures that the data collection process remains transparent, fair, and lawful~\cite{voigt2017eu}.
    \item Purpose limitation: the collected data is strictly used for the predefined purposes outlined in the consent forms. No data is processed for purposes beyond what is necessary for the execution of the Serenade project.
    \item Data minimization: Only the minimum amount of personal data required to achieve the project’s objectives is collected and processed by each data processor as mentioned in the previous section. 
    \item Retention limitation: personal data is stored only for the duration of the project, after which it is anonymized or deleted, in accordance with retention policies. Sensitive data retained on the home gateway is kept for a shorter retention period to reduce privacy risks. 
    \item Accountability: the system ensures accountability by integrating different techniques 
    which provide reliable logging and access control features to log all authentication events and all activities in the database. 
\end{itemize}

As AI methods are used for data analysis, a question may arise about compliance with regulations limiting the use of such systems (e.g., with the EU AI Act\footnote{https://artificialintelligenceact.eu/}). The system processes data that may reveal health conditions, and hence it is categorized as a high-risk AI system. However, no automatic decision nor diagnosis is provided by the system, since it only offers decision support for clinicians that have complete control and responsibility for any diagnosis or intervention. This mainly motivates the system compliance with the recent guidelines on these methods.


\section{Implemented Application}

The system is implemented as a Docker-based platform composed of five containers: frontend, backend, DBMS, identity and access management,
and application proxy.

The frontend is developed in Next.js and it serves as the user interface for all involved entities. Each entity is offered a personalized Web app for logging into the system and performing its tasks. The frontend communicates with the backend via REST APIs 
and authentication is implemented 
leveraging Keycloak. 

The backend ensures secure data processing by incorporating multiple layers of protection:
Since pseudonymization is a critical part of the solution, each subject's pseudonym (PID) is generated by using AES-256 encryption. For user-friendliness in the interfaces, a much shorter string (a proper name) is generated as an alias of the PID, ensuring no clashes between the strings occur. 
All communications are secured using HTTPS and 
token-based authentication to prevent interception and tampering. User authentication and access control are managed through Keycloak\footnote{https://www.keycloak.org/} providing fine-grained authorization and role-based access control to backend resources and APIs.
The association between PID and identifying data is also protected "at rest" by storing it in encrypted database tables. PostgreSQL is used as DBMS extended with Percona Transparent Data Encryption enabling on-the-fly encryption and decryption of data in selected tables.
%

Figure~\ref{fig:application} shows examples of the 
user interface as seen by different entities.
For example, HOS sees a list of subjects with their actual names and has a button to obtain and edit details about the patient and another button to see a dashboard with data analysis results. The IIT can see the open ticket related to a subject and has a chat panel to handle ticket details and communicate with the IMT.
The IMT has a list of installations but without data identifying the subjects and also has the chat panel for interacting with IIT.


\color{black}

\section{Conclusion and future work}
This paper illustrates the data protection analysis conducted for the design and implementation of the system behind the SERENADE project pilot. The adopted data protection solutions certainly do not address all privacy threats and can be certainly enhanced and extended. 
For example, we decided to expose to the data analysts some data, like a draft of the apartment map with the positioning of sensors as prepared by the IIT, that in principle may be considered a \emph{quasi-identifier}. However, these decisions are taken after an accurate evaluation of the trade-off between utility and the actual impact of the privacy threat.
Considering internal adversaries, good practice includes providing specific privacy training and enforcing legal terms. Further technological enhancements may include protecting data "in use" by running certain algorithms in TEE architectures and extending the encryption of data at rest to all data. The approved protocol also includes acquiring feedback from subjects and all the involved entities at the end of the pilot, allowing us to consider usability aspects and possible trade-offs with privacy protection.

\section*{Acknowledgment}
This work was partly supported by the MUSA and SERICS projects under the NRRP MUR program funded by the EU-NGEU. 

\bibliographystyle{ieeetr}
\bibliography{references}

\end{document}